\documentclass[10pt, onecolumn, conference]{IEEEtran}

\usepackage[utf8]{inputenc}
\usepackage[T1]{fontenc}
\usepackage[english]{babel}

\usepackage{graphicx,url}
\usepackage{indentfirst}
\usepackage{float}
\usepackage{booktabs}
\usepackage{array}
\usepackage{tabularx}
\usepackage{makecell}
\usepackage{ragged2e}

\usepackage{amsmath, amssymb}

\usepackage{color,colortbl}
\usepackage{xcolor}

\usepackage{listings}

\usepackage{tikz}
\usetikzlibrary{arrows.meta, shapes.geometric, positioning}

\usepackage{verbatim}
\usepackage{textcomp}
\usepackage{xspace}

\usepackage{hyperref}
\hypersetup{
    unicode=true,
    breaklinks=true,
    colorlinks=true,
    linkcolor=black,
    citecolor=black,
    urlcolor=black
}

\usepackage{todonotes}

\title{On-Premise SLMs vs. Commercial LLMs: Prompt Engineering and Incident Classification in SOCs and CSIRTs}

\author{
\IEEEauthorblockN{
    Gefté Almeida\IEEEauthorrefmark{1},
    Marcio Pohlmann\IEEEauthorrefmark{1},
    Alex Severo\IEEEauthorrefmark{1},
    Diego Kreutz\IEEEauthorrefmark{1},
    Tiago Heinrich\IEEEauthorrefmark{2},
    Lourenço Pereira\IEEEauthorrefmark{3}
}
\IEEEauthorblockA{
    \IEEEauthorrefmark{1}AI Horizon Labs, Federal University of Pampa (UNIPAMPA)
}
\IEEEauthorblockA{
    \IEEEauthorrefmark{2}Max Planck Institute for Informatics (MPI)
}
\IEEEauthorblockA{
    \IEEEauthorrefmark{3}Instituto Tecnológico de Aeronáutica (ITA)
}
}

\begin{document}

\maketitle

\begin{abstract}
In this study, we evaluate open-source models for security incident classification, comparing them with proprietary models. We utilize a dataset of anonymized real incidents, categorized according to the NIST SP 800-61r3 taxonomy and processed using five prompt-engineering techniques (PHP, SHP, HTP, PRP, and ZSL). The results indicate that, although proprietary models still exhibit higher accuracy, locally deployed open-source models provide advantages in privacy, cost-effectiveness, and data sovereignty.
\end{abstract}


\section{Introduction} 
\label{sec_introducao}


According to CERT.br, Brazil reported over 516k security incidents in 2024 and more than 181k in the first half of 2025, underscoring a persistent upward trend that challenges SOCs and CSIRTs to manage high alert volumes efficiently \cite{certbr}.
To alleviate this overload, AI-driven solutions, particularly prompt-engineering techniques such as Progressive Hint Prompting (PHP), have demonstrated over 90\% accuracy with models like GPT-4o and Gemini 2 \cite{severo2025framework}. However, the use of commercial LLMs involves high costs, dependence on external providers, and privacy risks imposed by the General Personal Data Protection Law (LGPD). As an alternative, locally executed \textit{open-source} models offer greater control, privacy, and autonomy, while maintaining competitive performance on accessible \textit{hardware} and reducing operational costs.

Given these limitations and the need to explore open alternatives, we perform an empirical evaluation of \textit{open-source} models using a balanced sample of anonymized real incidents, applying the same prompt-engineering techniques (PHP, SHP, and HTP) and the NIST SP 800-61r3 taxonomy (see Table \ref{tab_incident-categorization-nist}), following the previous study conducted with online LLMs \cite{severo2025framework}. In the evaluation, two groups of models were compared (Table~\ref{tab:modelos_grupo1} and Table~\ref{tab:modelos_grupo2})\footnote{\url{https://github.com/AILabs4All/FrameworkPE/blob/main/paper/metadata_apendice.md}}: the first group composed of smaller models, close to \textit{Llama-3.1-FoundationAI-SecurityLLM-Base-8B} \cite{Kassianik2025FoundationSec}, trained specifically for the cybersecurity domain; and the second group composed of larger models, around 70B parameters, representing the most robust open variants within their respective families. Two models from each family were selected, enabling comparison across different architectures and scales within the evaluated set.

\section{State of the Art} 
\label{sec_trabalhos_relacionados}

The efficacy of Large Language Models (LLMs) in classifying security incidents using structured taxonomies, such as NIST SP 800-61r3, relies heavily on domain-specific knowledge and contextual reasoning \cite{Salahuddin2025LessData, Kassianik2025FoundationSec}. General-purpose models often fail to capture cybersecurity-specific nuances, which motivates the adoption of \textit{Domain-Adaptive Continuous Pretraining} (DAP) methodologies. Examples such as \textit{Foundation-Sec-8B} demonstrate that continuous pretraining on specialized corpora yields substantial gains in benchmarks like CTIBench, while also improving typical SOC-related tasks including triage, alert enrichment, and TTP extraction \cite{Kassianik2025FoundationSec, Tellache2024Advancing}.

\begin{table}[!htp]
\caption{Security Incident Categorization based on NIST SP 800-61r3}
\centering
\scriptsize
\rowcolors{2}{lightgray}{white}
\label{tab_incident-categorization-nist}
\renewcommand{\arraystretch}{1.3}
\resizebox{1\textwidth}{!}{
\begin{tabular}{llll}
\hline
\textbf{Code} & \textbf{Category} & \textbf{Description} & \textbf{Priority} \\
\hline
CAT1 & Account Compromise & Unauthorized access to user or administrator accounts. & 5 \\
CAT2 & Malware & Infection by malicious code compromising devices or data. & 5 \\
CAT3 & Denial-of-Service Attack (DoS/DDoS) & Making systems or networks unavailable. & 4 \\
CAT4 & Data Exfiltration or Leakage & Unauthorized access, copying, or disclosure of sensitive data. & 5 \\
CAT5 & Vulnerability Exploitation & Use of known or unknown flaws to compromise assets. & 5 \\
CAT6 & Insider Abuse & Intentional or negligent actions by internal users. & 5 \\
CAT7 & Social Engineering & Deceiving people to obtain access or information. & 3 \\
CAT8 & Physical or Infrastructure Incident & Physical breach impacting computational assets. & 4 \\
CAT9 & Unauthorized Modification & Unauthorized changes to systems, data, or configurations. & 3 \\
CAT10 & Misuse of Resources & Unauthorized use of systems for other purposes. & 2 \\
CAT11 & Vendor/Third-Party Problem & Incident originating from a third-party security failure. & 4 \\
CAT12 & Intrusion Attempt & Hostile attempts to break in not yet confirmed as successful. & 3 \\
\hline
\end{tabular}
}
\end{table}

In parallel, the study by Irugalbandara \textit{et al.} \cite{irugalbandara2024slam} evaluated nine \textit{open-source} models and twenty-nine quantized variants, showing that self-hosted solutions can reduce costs by factors ranging from 5$\times$ to 29$\times$, while maintaining latency equal to or lower than proprietary models and offering greater operational stability. Whereas GPT-4 operates at an approximate cost of US\$\,0.09 per thousand \textit{tokens}, local models range between US\$\,0.003 and US\$\,0.018 per thousand \textit{tokens}, reinforcing the feasibility of large-scale adoption.

These findings contrast directly with the use of proprietary LLMs (such as GPT-4o and Claude), whose dependency on external providers introduces risks related to privacy, legal compliance (e.g., LGPD), and data sovereignty \cite{Pan2025CostBenefit, Noreika2025OpenSource}. Conversely, \textit{open-source} models (such as \textit{Llama}, \textit{Mistral}, and \textit{Qwen}) offer greater operational control, transparency, and flexibility for fine-tuning capabilities that are essential in environments handling sensitive information. Total Cost of Ownership (TCO) analyses indicate that \textit{on-premise} deployments become economically advantageous after short break-even periods, particularly for small and medium-sized models \cite{Pan2025CostBenefit}. Thus, open-source models emerge as viable and strategically attractive alternatives for corporate and institutional scenarios requiring high levels of privacy, autonomy, and auditability.



\section{\textit{Pipeline} for Automated Classification with SLMs}
\label{sec_metodologia}

\begin{figure}[!ht]
    \centering
    \includegraphics[width=1\linewidth]{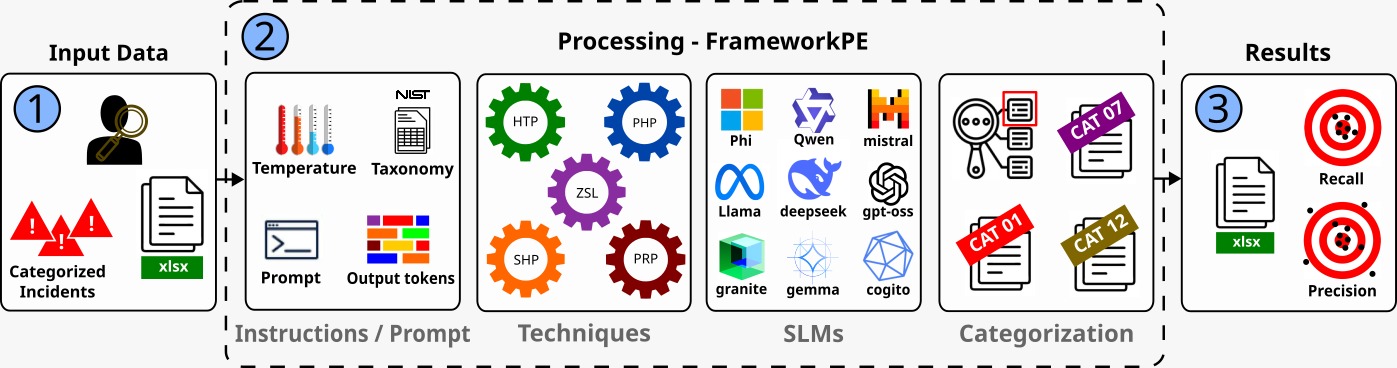}
    \caption{Flowchart of the automated classification pipeline using SLMs.}
    \label{fig_fluxo_eng_prompt}
    \vspace{0.3em}
\end{figure}

The methodological architecture adopted in this work is based on a modular automated classification pipeline composed of five stages: Input Data, Preprocessing, Processing, Analysis, and Output, as illustrated in \autoref{fig_fluxo_eng_prompt}. This structure is inspired by the framework introduced in previous work \cite{severo2025framework}, but it has been fully adapted for local execution, enabling the assessment of \textit{open-source} models in real-world scenarios of security incident categorization.

\textbf{Input Data}. The incident reports used in this study are the same as those employed in the original experiment with LLMs. The dataset consists of a balanced sample of 24 security incidents selected from the 194 records used in \cite{severo2025framework}. All records were previously anonymized using the AnonLFI tool\footnote{\url{https://github.com/gt-rnp-lfi/anon}}, ensuring the removal of names, IP addresses, and other sensitive identifiers. The reference labels (\textit{ground truth}) were preserved exactly as defined by two cybersecurity specialists in the prior study, serving as the baseline for comparative evaluation of the open-source models.

\textbf{Preprocessing}. The preprocessing stage maintained the same semantic structure adopted in the original pipeline, but was adapted to the formats required by \textit{open-source} models. Input and output \textit{token} limits were defined, and a uniform temperature was applied across all inferences. The entire process was automated through Python scripts, ensuring reproducibility and traceability across runs.

\textbf{Processing}. 
The processing stage constitutes the core of the pipeline and is responsible for applying the five prompt-engineering techniques (PHP, SHP, HTP, PRP, and ZSL) to different models executed locally via the Ollama \textit{framework} in a controlled environment with GPU support. Inference parameters such as temperature, maximum number of \textit{tokens}, and iteration depth were standardized to ensure comparability across models, while metrics such as average execution time, memory usage, and response length were monitored to assess efficiency and accuracy. The models, ranging from 7B to 70B parameters, span distinct \textit{open-source} families and architectures, incorporating varied attention mechanisms, activation functions, and normalization strategies. Additional architectural details are provided in Tables \ref{tab:modelos_grupo1} and \ref{tab:modelos_grupo2}.

\begin{table}[!htp]
\caption{Group 1 models and their main architectural characteristics.}
\centering
\scriptsize
\rowcolors{2}{lightgray}{white}
\label{tab:modelos_grupo1}
\renewcommand{\arraystretch}{1.3}
\resizebox{0.8\textwidth}{!}{
\begin{tabular}{lccccc}
\hline
\textbf{Model} & \textbf{Size} & \textbf{Context Window (tokens)} & \textbf{Attention} & \textbf{Activation} & \textbf{Normalization} \\
\hline
Cogito 70B & 70B & 128,000 & MHA & SwiGLU & RMSNorm \\
DeepSeek R1 70B & 70B & 128,000 & MLA & SwiGLU & RMSNorm \\
Falcon3 10B & 10B & 32,000 & GQA & ReLU & LayerNorm \\
Gemma 2 27B & 27B & 8,192 & GQA & GeGLU & RMSNorm \\
Gemma 3 27B & 27B & 128,000 & GQA & GeGLU & RMSNorm \\
GPT-OSS 20B & 20B & 131,072 & MoE & SwiGLU & RMSNorm \\
Llama 3.3 70B & 70B & 131,072 & GQA & SwiGLU & RMSNorm \\
Mistral Small 24B & 24B & 32,768 & GQA & SwiGLU & RMSNorm \\
Phi-4 14B & 14B & 16,000 & MHA & GeGLU & RMSNorm \\
Qwen2.5 32B & 32B & 131,072 & GQA & SwiGLU & RMSNorm \\
Qwen3 32B & 32B & 131,072 & GQA & SwiGLU & RMSNorm \\
\hline
\end{tabular}
}
\end{table}

\begin{table}[!htp]
\caption{Group 2 Models and their main architectural characteristics.}
\centering
\scriptsize
\rowcolors{2}{lightgray}{white}
\label{tab:modelos_grupo2}
\renewcommand{\arraystretch}{1.3}
\resizebox{0.8\textwidth}{!}{
\begin{tabular}{lccccc}
\hline
\textbf{Model} & \textbf{Size} & \textbf{Context Window (tokens)} & \textbf{Attention} & \textbf{Activation} & \textbf{Normalization} \\
\hline
Qwen3 8B & 8B & 262{,}144 & GQA & SwiGLU & RMSNorm \\
Qwen2.5 7B & 7B & 131{,}072 & GQA & SwiGLU & RMSNorm \\
Cogito 8B & 8B & 128{,}000 & MHA & SwiGLU & RMSNorm \\
DeepSeek R1 8B & 8B & 128{,}000 & MLA & SwiGLU & RMSNorm \\
Falcon3 7B & 7B & 32{,}000 & GQA & ReLU & LayerNorm \\
Gemma 2 9B & 9B & 8{,}192 & GQA & GeGLU & RMSNorm \\
Gemma 3 12B & 12B & 128{,}000 & GQA & GeGLU & RMSNorm \\
Granite3.2 8B & 8B & 131{,}072 & MHA & SwiGLU & RMSNorm \\
Llama 3.1 8B & 8B & 128{,}000 & MHA & SwiGLU & RMSNorm \\
Mistral 7B & 7B & 32{,}768 & GQA, SWA & SwiGLU & RMSNorm \\
Foundation-Sec 8B & 8B & 128{,}000 & MHA & SwiGLU & RMSNorm \\
\hline
\end{tabular}
}
\end{table}



\section{Results and Discussion} 
\label{sec_resultados}

The results derived from the models were benchmarked against the baseline established in the prior study \cite{severo2025framework}, which utilized proprietary models, evaluating both accuracy and operational efficiency. The evaluation involved two primary metrics: classification accuracy, measured by the correspondence between the category assigned by the model and the expert-defined \textit{ground truth} and computational efficiency, assessed through average inference time and GPU usage.



\subsection{Comparison Across Models and Techniques}
\label{sec:analise_tecnicas}

\begin{figure}[!ht]
    \centering
    \includegraphics[width=\linewidth]{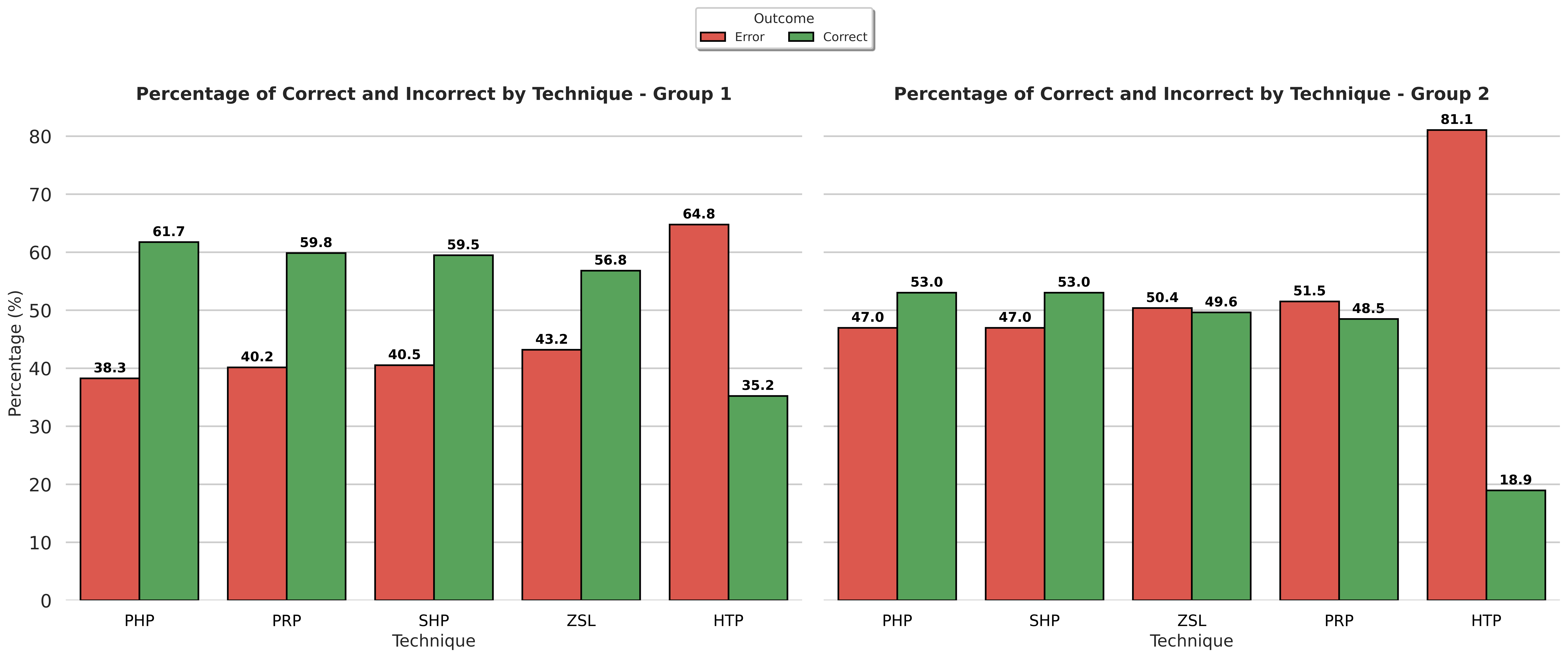}
    \caption{Percentage of correct and incorrect predictions by prompt technique.}
    \label{fig:percentual_tecnica}
\end{figure}

\begin{figure}[!ht]
    \centering
    \includegraphics[width=\linewidth]{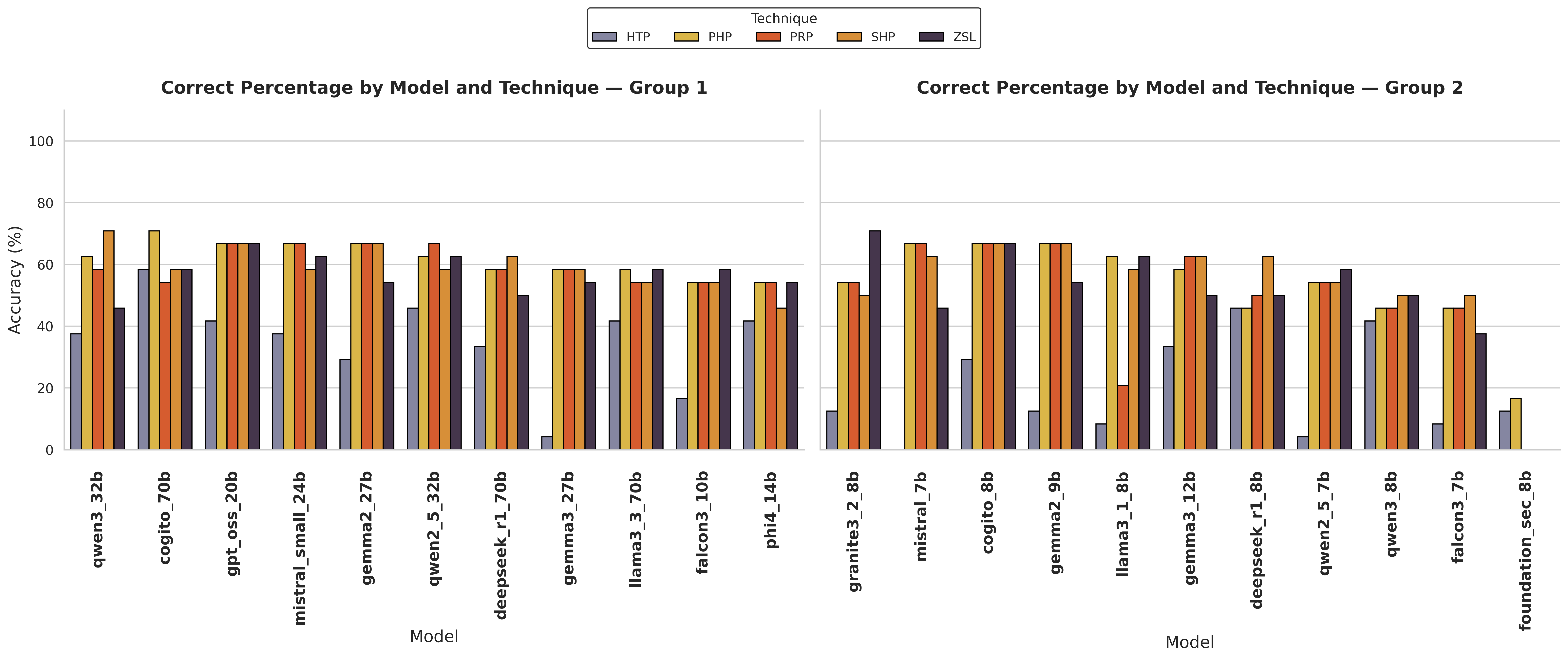}
    \caption{Percentage of correct and incorrect predictions by Model × Prompt Technique.}
    \label{fig:percentual_modelo_tecnica}
\end{figure}

Figures \ref{fig:percentual_tecnica} and \ref{fig:percentual_modelo_tecnica} present a comparison of the performance of the five prompt-engineering strategies (PHP, SHP, PRP, HTP, and ZSL) applied to both groups of models. It is evident that the approaches based on progressive hints (PHP and SHP) continue to exhibit the best overall performance, achieving higher and more consistent accuracy rates across the evaluated models. This result aligns with the pattern identified in \cite{severo2025framework} and reinforces the notion that incremental hinting facilitates semantic alignment between the model and the categorization instructions, contributing to improved contextual generalization.

In Group 1, composed predominantly of large models (between 20B and 70B parameters), \textit{Progressive Hint Prompting} (PHP) achieved the best average performance, with an accuracy of 61.7\%, followed by PRP and SHP, both in the 59–60\% range. \textit{Zero-Shot Learning} (ZSL) obtained an intermediate result (56.8\%), whereas \textit{Hypothesis Testing Prompting} (HTP) exhibited significantly lower performance (35.2\%). The gap of approximately 26 percentage points between PHP and HTP highlights the importance of iterative refinement and the gradual decomposition of reasoning in higher-capacity models. Additionally, architectures featuring optimized attention mechanisms (such as GQA and MoE) and modern activation functions (SwiGLU, GeGLU) showed better responsiveness to progressive \textit{prompting} techniques.

In Group 2, composed of smaller models (between 7B and 12B parameters), a general reduction in accuracy levels was observed, although the superior performance pattern of PHP and SHP persisted, with both achieving averages close to 53\%. PRP and ZSL displayed similar performance (49–51\%), while HTP was again the least effective method, reaching only 18.9\% accuracy. This more pronounced decline in techniques requiring complex reasoning indicates that smaller-scale models are more sensitive to the cognitive load of the instructions and degrade more rapidly when the \textit{prompt} demands hypothetical or chained inferences.

In summary, the comparison between the groups shows that both model size and attention architecture directly influence the stability of \textit{prompting} techniques. Models with optimized attention mechanisms, such as GQA and MoE, as well as those with larger context windows, exhibited greater consistency across the evaluated techniques, whereas models relying on conventional attention (MHA) or simpler activation functions (ReLU) showed higher performance variability. These findings indicate that architectural \textit{design} and the choice of \textit{prompting} strategy interact in a decisive manner in determining overall results, with progressive approaches (PHP and SHP) providing the greatest robustness even under different model configurations.

\section{Final Considerations and Future Work} 
\label{sec_consid_finais_trab_fut}

This study indicates that SLMs are a technically viable alternative for automated classification of security incidents, particularly in institutional environments that require data privacy, cost predictability, and operational autonomy. Although open-source models between 8B and 20B parameters reached accuracies around 60\%, below the levels above 90\% typically achieved by LLMs, they demonstrated estabilidade and semantic consistency for initial triage and decision-support tasks in SOCs and CSIRTs. Prompting strategies such as \textit{Progressive-Hint Prompting} and \textit{Self-Hint Prompting} mitigated context limitations inherent to smaller models, while the use of the Ollama framework ensured full control over data, compliance with the LGPD, and predictable operational costs.

For future work, we plan to extend the dataset with incidents from multiple sectors and languages, and evaluate \textit{fine-tuning} techniques such as LoRA to improve model performance in specific cybersecurity domains. We also intend to incorporate more granular metrics, including \textit{precision}, \textit{recall} and \textit{F1-score}, implement a continuous-learning pipeline to reassess models as new incidents arise, and develop a visual interface for SOC and CSIRT workflows with emphasis on explainability and traceability. These advances aim to consolidate SLMs as autonomous, interpretable and auditable solutions, strengthening digital sovereignty and cyber intelligence in critical environments.

\bibliographystyle{IEEEtran}
\bibliography{refs}

\end{document}